# The Impact of Natural Disasters on Economic Growth: A Comparative Analysis Between Developed and Developing Countries

Group H: Sharmistha Chakrabarti (sc5466), Md Shah Naoaj (mn3116), Yuanmeng Yang (yy4025), Shuman Zhang (sz3658), Xinru Chen (xc2115)

## I. Introduction and Background

Natural disasters are costly to the economy. They can affect fiscal balance directly and indirectly. To address post-disaster relief and recovery efforts, governments of affected countries face revenue losses and spending pressures. These costs are direct. The indirect costs consist of a wider macroeconomic impact, such as employment issues and output losses due to the inefficiency and destruction of factories and agriculture (Koetsier, 2017). Moreover, the severity and frequency of disasters also pose restrictions on the governments' access to financial markets, worsening their borrowing conditions. This could further exacerbate fiscal vulnerabilities and sovereign risks (Mallucci, 2020). This paper compares the impacts of severe natural disasters on growth in developed and developing countries around the world.

## II. Literature Review

Several past papers have conducted research on the relationship between natural disasters and economic growth. Generally, natural disasters have negative effects on economic growth. Klomp and Valckx (2014) verified the negative relationship between natural disasters and economic growth per capita. Moreover, negative effects induced by natural disasters to different countries may vary. Ikeda and Loayza (2013) found that developed countries suffer less from severe disasters than developing countries. Panwar and Sen(2019)'s results indicate that natural disasters have diverse economic impacts across economic sectors depending on disaster types and their intensity. After disasters, financial expenditures are needed to help recovery and disasters may influence government's activity as well. Nishizawa, Roger and Zhang (2019) separate financial buffers for natural disasters into two parts: immediate needs and borrowing of long-term recovery needs.

III. Data Description

We use annual natural disaster data from the EM-DAT database. We use four filter conditions to adjust the database as our project data: (i) the period is from 1990 to 2019; (ii) disaster subgroups are Hydrological and Meteorological. Data for government expenditure and government revenue are from the IMF World Economic Outlook database (Table 1).

In recent years, the number of total natural disasters have generally risen with some minor fluctuations. The loss of damages caused by natural disasters has also experienced great fluctuations, with the tendency to increase during the past 30 years. Compared to total disasters, severe disasters have no obvious trend and the number of severe disasters overall remains steady (Figure 1).

As Figure 2 shows, developed countries generally suffer more damage loss than developing countries but the total damage loss as percent of GDP in developing countries is much smaller than developed countries. Figure 3 shows impact by category of disaster. For four natural disaster types, droughts and floods influence more people than storms and extreme temperatures do. However, droughts have the least total damage of the four types so it may have a wider range impact but less fiscal cost (Figure 3).

IV. Methodology and Data

$Y_{i,t} = \beta_1 X_{i,t} + \beta_2$(population density, percent educated, inflation, initial output-GDP per capita)$+\varepsilon_{i,t}$

Given that we have panel data measuring disasters and economic outcomes for countries around the world over several years, we used a fixed effects panel regression to capture the country-specific characteristics that are not quantified in this model, e.g. political system, infrastructure capacity, or land area. We also used a Hausman test to decide between fixed or random effects, and used a pooled OLS model as a second approach. In our model, we use per capita GDP growth and agricultural growth as the dependent variables and the natural disaster dummy as the independent variable with some control variables. We conducted this analysis separately for developed and developing countries to see how disasters affect each group differently (Fomby et al.). Our initial expectation was that developing countries suffer more in terms of growth.

In our regression model, we include 2 main dependent variables: per capita GDP growth and agricultural growth. The independent variable is the natural disaster dummy. We assign severe disasters a value of 1 and assign a value of 0 to those that are not considered severe. We also take population density, percent with primary education, inflation and initial output-GDP per capita as our control variables.

We specifically include population density as a control since it affects the amount of money needed to rebuild after a disaster. If two countries are about the same size, the smaller the population density is, the smaller the impact of the disaster, and the smaller the cost of post-disaster reconstruction.

For the disaster dummy variable, following the example of a paper by Panwar and Sen, we define the severity of natural disaster as any disaster that is above the 75th percentile in the damage-to-GDP ratio and that the number of people affected (including killed)-to-total population ratio is greater than 0.01%. This allows us to exclude the disaster that had a minimal effect either because the area affected was not well-populated or the disaster itself lacked intensity.

To distinguish between developed and developing countries, we take the countries whose HDI (the Human Development Index) scores are greater than or equal to 0.75 as "developed." We chose to label the countries as "developed" if they reached developed status at some point over the 30 year time span. That would imply that even if they were not above .75 for the entire time period, they were likely fairly close to the cutoff given the pace at which countries develop.

V. Results

As mentioned above our data was in panel form so we focused on two main models accordingly: fixed effects and pooled OLS. The fixed effects model was expected to be effective since it would account for the average differences across countries, whether those differences were observable or unobservable. However, it was still necessary to ensure that a fixed effects model was more appropriate than a random effects model so we ran a Hasuman test to ensure that was the case. The p-value was close to 0 at $2.2e^{-16}$, so we accordingly rejected the null hypothesis that the random effects model is consistent and moved forward with a fixed effects model.

It was also necessary to check for heteroskedasticity so we conducted a Breusch-Pagan test. The p-value for this test was also very small at $6.571e^{-9}$ so it was clear that the residuals were heteroskedastic. In order to mitigate that issue, we used robust standard errors in our regression. There are several different options for type of robust standard error, but since a comparison of heteroskedasticity-error 0, 1, 2, and 3 showed little difference and we had a large number of observations, we used HC2 in our main regressions.

We checked for multicollinearity among our chosen variables as well. Initially, one variable, private sector credit, did have a strong correlation with a few others, such as agricultural growth and GDP per capita. However, private sector credit was not essential to our model specification and we consequently decided to remove it from our specification and move forward with our remaining control variables. As you can see from Table 2, the correlation between other variables could be categorized as either moderate or weak. Therefore, we were able to continue with our regression analysis.

Table 3 shows the overall regression using the entire collection of countries, both developed and developing. The coefficient on disaster intensity shows the effect of a severe disaster on agricultural growth and was significant at the 1% level, which means a severe disaster was associated with a decline in agricultural growth. This result matched our initial expectations which were informed by the idea that severe natural disasters would likely affect agricultural endeavors and therefore, affect agricultural growth in general.

We also examined the differential effect of disasters on developed and developing countries by running a separate regression for each of those groups, the results of which can be found in Table 4. The coefficients for the fixed effects model were insignificant, but we did get the expected result in terms of comparison between developed and developing countries. We anticipated that severe disasters would cause a more adverse effect on developing countries than developed countries. The signs of the coefficients support this expectation since the coefficients for developing countries were negative, and the coefficients for developed countries were not.

Since the fixed effects model did not return any significant results we also employed a pooled OLS model to better compare the effect on developed and developing countries.

From the results of the pooled OLS shown in Table 5, we can see that our main explanatory variables are statistically significant for both developing and developed countries with the expected negative sign for the disaster dummy variable. The F-test shows that the model itself is the best fitted model. From the pooled OLS we observe that developing countries' agricultural growth impacted twice as much severely as the agricultural growth of developed countries, and while per capita GDP growth of developing countries is impacted severely, the developed countries do not have any statistical impact.

The $R^2$ for all results in these models were fairly low, and in the future, we would likely attempt to add in more relevant variables to make the model a better fit since our current model could be suffering from omitted variable bias.

## VI. Findings and Policy Implications

This study re-examines the impact of natural disasters on economic growth in the perspective of developed and developing countries. Based on panel data consisting of developing and developed countries over the period 1990-2019 and using different statistical techniques we find that– i) Developing countries' agricultural growth impacted twice as much severely as the agricultural growth of developed countries. ii) While per capita GDP growth of developing countries is impacted severely, the developed countries don't have any statistical impact. The study confirms the finding of previous studies such as Panwar and Sen (2019) , and Nishizawa et al (2019) that the economic impact of natural disasters are statistically stronger in developing countries.

These findings may stimulate the policymakers especially in developing countries to explore the efficacy of viable ex-ante disaster risk financing tools (such as insurance, micro-insurance and catastrophic bonds). This would not only safeguard population and physical assets but also ensure adherence to the sustainable development goals. Future research can be done on estimating the fiscal buffers to combat the natural disaster, and identify the causality between climate change and natural disaster and possible mitigants.

It may also be beneficial to use a less stringent definition for severe disaster. We used disasters that passed two tests relating to population affected and damage to GDP. One possible change moving forward would be to relax that definition by using only one of those two metrics and potentially capture the effect of a different level of disaster.

## VII. References

Fomby, T., Y. Ikeda, and N. Loayza (2013), The Growth aftermath of Natural Disasters. *Journal of Applied Econometrics*, 28(3), 412–434.

Klomp, J., and K. Valckx (2014), Natural disasters and economic growth: A meta-analysis. *Global Environmental Change*, 26, 183-195.

Koetsier, I. (2017). Natural disasters and (future) government debt. *Public or Private Goods?*, 48–74.

Mallucci, E. (2020). Natural disasters, climate change, and sovereign risk. *FEDS Notes*, *2020*(2813).

Nishizawa, H., Roger, S., & Zhang, H. (2019). Fiscal buffers for natural disasters in Pacific Island countries. *SSRN Electronic Journal*.

Naoaj, M.S. (2019). Philippines: Vulnerability to Climate Change and Natural Disaster. Selected Issues Paper, International Monetary Fund, Washington, DC.

Naoaj, M. S. (2023). From Catastrophe to Recovery: The Impact of Natural Disasters on Economic Growth in Developed and Developing Countries. European Journal of Development Studies, 3(2), 17–22. https://doi.org/10.24018/ejdevelop.2023.3.2.237

Panwar, Vikrant, and Subir Sen. "Economic Impact of Natural Disasters: An Empirical Re-Examination." *Margin: The Journal of Applied Economic Research*, vol. 13, no. 1, 2019, pp. 109–139.

Torres-Reyna , Oscar (2007). Panel Data Analysis Fixed and Random Effects Using Stata. https://www.princeton.edu/~otorres/Panel101R.pdf.

VIII. Tables and Figures

**Figure1: Trends in Damages from Natural Disasters Worldwide, 1990–2019
(left:total disasters right: severe disasters)**

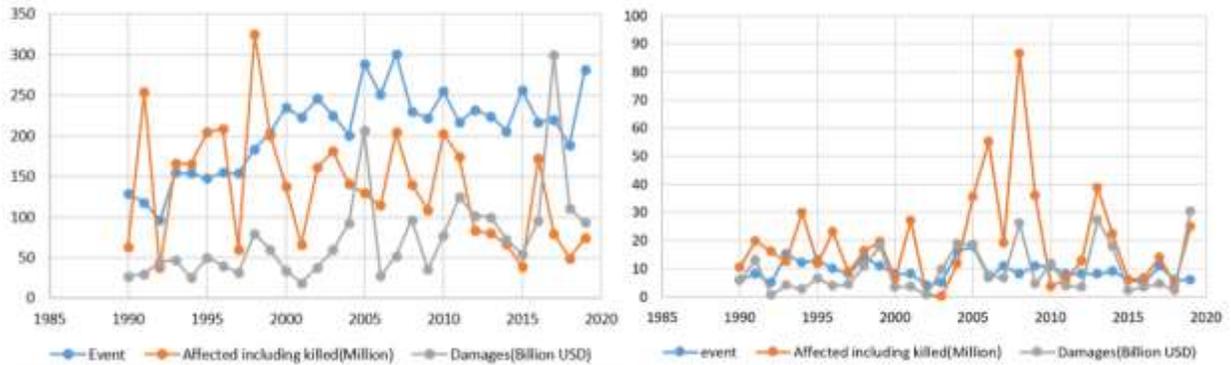

Source: Authors' calculation using data from Emergency Events Database (EM-DAT).

**Figure 2:** Average Damages from Natural Disasters by Country groups, 1990–2019
(left:total disasters right: severe disasters )

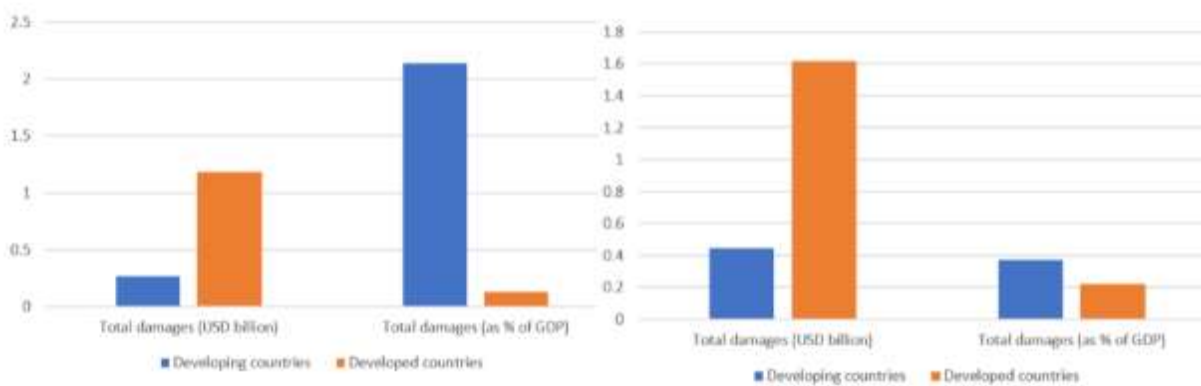

Source: Authors' calculation using data from Emergency Events Database (EM-DAT).

**Figure 3:** Average Damage per Disaster, by type (1990-2019)

|  | Average population Affected including killed(thousands of person) | | Total Damage (USD thousands) | |
| --- | --- | --- | --- | --- |
| Storm | 387.75 | 413.30 | 625.63 | 993.92 |
|  |  | 401.56 |  | 111.38 |
| Flood | 1004.94 | 1609.70 | 245.96 | 473.87 |
|  |  | 632.52 |  | 86.85 |
| Extreme temperature | 243.07 | 338.73 | 124.10 | 177.90 |
|  |  | 16.51 |  | 8.07 |
| Drought | 4215.03 | 5758.75 | 122.10 | 727.50 |
|  |  | 4442.56 |  | 41.54 |

\* blue for developed countries and yellow for developing countries
Source: Authors' calculation using data from Emergency Events Database (EM-DAT).

Table 1: Descriptive Statistics

| Statistic | per_capita_GDP_growth | agricultural.growth | population.density | Initial.output.GDP.per.capita | Terms.of.Trade.adjustment |
|---|---|---|---|---|---|
| N | 6,022 | 5,845 | 6,130 | 6,030 | 5,127 |
| Mean | 2.979 | 13.073 | 184.927 | 10,201.560 | -5,395,589,421,456.000 |
| St. Dev. | 3.952 | 10.404 | 430.850 | 15,544.950 | 86,674,170,676,955.000 |
| Min | -47.591 | 0.053 | 1.406 | 90.532 | -1,400,000,000,000,000.000 |
| Pctl(25) | 1.019 | 3.933 | 35.542 | 902.070 | -73,000,000,000.000 |
| Pctl(75) | 5.226 | 19.581 | 238.391 | 10,016.570 | 33,400,000,000.000 |
| Max | 23.999 | 65.598 | 20,213.570 | 100,600.600 | 484,000,000,000,000.000 |

| Statistic | Percent.Educated..primary. | Economic.Openness.Stocks.traded..total.value....of.GDP. | Private.Sector.Credit |
|---|---|---|---|
| N | 4,779 | 4,063 | 5,260 |
| Mean | 0.112 | 54.798 | 65.789 |
| St. Dev. | 0.044 | 71.609 | 56.058 |
| Min | 0.033 | 0.001 | 0.000 |
| Pctl(25) | 0.078 | 7.175 | 22.164 |
| Pctl(75) | 0.142 | 73.129 | 109.158 |
| Max | 0.303 | 715.166 | 217.761 |

Table 2: Correlations between variables used in the model after dropping Private Sector Credit.

| | Year | per_capita_GDP_growth | ag_growth | disaster intensity | population density | Initial output-GDP per capita | Developing Status | Primary Education | Percent Educated (primary) | Inflations |
|---|---|---|---|---|---|---|---|---|---|---|
| Year | 1 | | | | | | | | | |
| per_capita_GDP_growth | .01 | 1 | | | | | | | | |
| ag_growth | -.19 | .09 | 1 | | | | | | | |
| disaster intensity | -.06 | .02 | -.01 | 1 | | | | | | |
| population density | -.01 | .08 | .10 | .01 | 1 | | | | | |
| Initial output-GDP per capita | .16 | -.19 | -.61 | -.05 | -.01 | 1 | | | | |
| Developing Status | -.06 | .13 | -.67 | .00 | -.22 | .51 | 1 | | | |
| Primary Education | .04 | .40 | .10 | .03 | .07 | -.18 | .03 | 1 | | |
| Percent Educated (primary) | -.15 | -.09 | .51 | .05 | -.08 | -.48 | -.52 | -.13 | 1 | |
| Inflations | -.07 | -.10 | .06 | -.01 | -.02 | -.03 | .00 | -.02 | .03 | 1 |

Table 3: General results of effect on agricultural growth using the entire selection of countries.

|  | Agricultural Growth |
|---|---|
| disaster intensity | -3.335*** |
|  | (0.503) |
| population density | 0.002* |
|  | (0.001) |
| Percent Educated (primary) | 73.702*** |
|  | (7.223) |
| Inflation | 0.001 |
|  | (0.003) |
| Initial output-GDP per capita | 0.000*** |
|  | (0.000) |
| Num.Obs. | 4598 |
| R2 | 0.401 |
| R2 Adj. | 0.375 |
| Std.Errors | Robust |

Table 3: General results of effect on agricultural growth using the entire selection of countries.

| | Developing GDP Growth | Developing Agricultural Growth | Developed GDP Growth | Developed Agricultural Growth |
|---|---|---|---|---|
| disaster intensity | -0.745 | -0.236 | 0.195 | 0.648* |
| | (0.502) | (1.025) | (0.333) | (0.282) |
| population density | 0.002* | -0.009*** | 0.006*** | 0.002 |
| | (0.001) | (0.001) | (0.001) | (0.002) |
| Percent Educated | -2.594 | -18.626** | -27.662*** | -20.001* |
| | (5.868) | (6.984) | (7.168) | (7.772) |
| Inflations | -0.001 | 0.001*** | -0.003** | 0.001 |
| | (0.000) | (0.000) | (0.001) | (0.002) |
| Initial output-GDP per capita | -0.001 | -0.007*** | 0.000*** | 0.000*** |
| | (0.000) | (0.001) | (0.000) | (0.000) |
| Num.Obs. | 1125 | 1107 | 1272 | 1234 |
| R2 | 0.048 | 0.465 | 0.250 | 0.446 |
| R2 Adj. | -0.009 | 0.433 | 0.205 | 0.413 |

Table 5: Pooled OLS regression results comparing the effect of severe disasters on growth for developing and developed countries.

| | (1) | (2) | (3) | (4) |
|---|---|---|---|---|
| | Impact on Agri Growth | | Impact on Per Capita GDP Growth | |
| VARIABLES | developing | developed | developing | developed |
| d_dummy_1 | -4.970*** | -2.568*** | -45.81** | -10.14 |
| | (0.878) | (0.744) | (23.49) | (24.63) |
| p_density | 0.000866* | 0.000289 | -0.0109 | -0.0252*** |
| | (0.000512) | (0.000224) | (0.0135) | (0.00746) |
| edu | 2.650 | 115.8*** | -1,186*** | -422.1*** |
| | (7.806) | (3.256) | (207.7) | (108.1) |
| inflation | -0.000676 | 0.00229*** | 0.0967*** | 0.0868*** |
| | (0.000725) | (0.000357) | (0.0192) | (0.0119) |
| output | 0.00154*** | -0.00106*** | -0.0331*** | -0.00656 |
| | (0.000378) | (0.000214) | (0.0101) | (0.00712) |
| Constant | 19.50*** | -1.578*** | 715.6*** | 600.2*** |
| | (1.590) | (0.518) | (42.04) | (17.31) |
| Observations | 1,107 | 3,491 | 1,125 | 3,554 |
| Prob>F | 0.00*** | 0.00*** | 0.00*** | 0.00*** |
| R-squared | 0.047 | 0.309 | 0.064 | 0.020 |